# Cataloguing MoSi$_2$N$_4$ and WSi$_2$N$_4$ van der Waals Heterostructures: An Exceptional Material Platform for Excitonic Solar Cell Applications

Che Chen Tho, Chenjiang Yu, Qin Tang, Qianqian Wang, Tong Su, Zhuoer Feng, Qingyun Wu, C. V. Nguyen, Wee-Liat Ong, Shi-Jun Liang, San-Dong Guo, Liemao Cao, Shengli Zhang, Shengyuan A. Yang, Lay Kee Ang, Guangzhao Wang*, Yee Sin Ang*

Che Chen Tho, Qianqian Wang, Tong Su, Zhuoer Feng, Dr. Qingyun Wu, Prof. Shengyuan A. Yang, Prof. Lay Kee Ang, Dr. Yee Sin Ang*
Science Mathematics and Technology Cluster, Singapore University of Technology and Design, 487372, Singapore
Email: yeesin_ang@sutd.edu.sg

Chenjiang Yu, Qing Tang, Prof. Guangzhao Wang*
School of Electronic Information Engineering, Key Laboratory of Extraordinary Bond Engineering and Advanced Materials Technology of Chongqing, Yangtze Normal University, Chongqing, China, 408100
E-mail: wangyan6930@yznu.edu.cn

Prof. C. V. Nguyen
Department of Materials Science and Engineering, Le Quy Don Technical University, Ha Noi, Vietnam, 100000

Prof. Wee-Liat Ong
ZJU-UIUC Institute, College of Energy Engineering, Zhejiang University, Jiaxing, Haining, Zhejiang, China, 314400
State Key Laboratory of Clean Energy Utilization, Zhejiang University, Hangzhou, Zhejiang, China, 310027

Prof. Shi-Jun Liang
Institute of Brain-Inspired Intelligence, National Laboratory of Solid State Microstructures, School of Physics, Collaborative Innovation Center of Advanced Microstructures, Nanjing University, Nanjing, China, 210093

Prof. San-Dong Guo

School of Electronic Engineering, Xi'an University of Posts and Telecommunications, Xi'an, China, 710121

Prof. Liemao Cao

College of Physics and Electronic Engineering, Hengyang Normal University, Hengyang, China, 421002

Prof. Shengli Zhang

MIIT Key Laboratory of Advanced Display Materials and Devices, School of Materials Science and Engineering, Nanjing University of Science and Technology, Nanjing, China, 210094

Keywords: density functional theory simulations, van der Waals heterostructures, 2D materials, excitonic solar cells, electrical contacts

**Abstract**

Two-dimensional (2D) materials van der Waals heterostructures (vdWHs) provides a revolutionary route towards high-performance solar energy conversion devices beyond the conventional silicon-based *pn* junction solar cells. Despite tremendous research progress accomplished in recent years, the searches of vdWHs with exceptional excitonic solar cell conversion efficiency and optical properties remain an open theoretical and experimental quest. Here we show that the vdWH family composed of $MoSi_2N_4$ and $WSi_2N_4$ monolayers provides a compelling material platform for developing high-performance ultrathin excitonic solar cells and photonics devices. Using first-principle calculations, we construct and classify 51 types of $MoSi_2N_4$ and $WSi_2N_4$-based $[(Mo,W)Si_2N_4]$ vdWHs composed of various metallic, semimetallic, semiconducting, insulating and topological 2D materials. Intriguingly, $MoSi_2N_4$/(InSe, $WSe_2$) are identified as Type-II vdWHs with exceptional excitonic solar cell power conversion efficiency reaching well over 20%, which are competitive to state-of-art silicon solar cells. The $(Mo,W)Si_2N_4$ vdWH family exhibits strong optical absorption in both the visible and ultraviolet regimes. Exceedingly large peak ultraviolet absorptions over 40%, approaching the maximum absorption limit of a free-standing 2D material, can be achieved in $(Mo,W)Si_2N_4/\alpha_2$-$(Mo,W)Ge_2P_4$ vdWHs. Our findings unravel the enormous potential of $(Mo,W)Si_2N_4$ vdWHs in designing ultimately compact excitonic solar cell device technology.

# 1. Introduction

Solar-to-electricity conversion offers a promising clean energy source to power a more sustainable technology landscape in the future. Despite the widespread deployment of conventional silicon-based solar cell, the continual improvement of the solar-to-electricity conversion efficiency remains a formidable challenge. Conventional crystalline silicon *pn* junction solar cell is plagued by nonradiative recombination of photogenerated electron-hole pairs,[1] which severely limits its solar-to-electricity conversion efficiency.[2] Silicon solar cells also entails the use of thicker layers to achieve longer optical path for better light absorption, which fundamentally limits their deployment in ultracompact and low-mass solar cell design.[3] The search of novel nanomaterials and device architectures beyond the conventional silicon *pn* junction solar cell remain an open research challenge that urgently needs to be addressed.

Excitonic solar cell has recently emerged as a compelling candidate for high-efficiency solar-to-electricity conversion. In contrast to conventional semiconductor-based *pn* junction solar cell where light-to-electricity conversion occurs exclusively within the same bulk material, excitonic solar energy conversion is an interfacial effect that arises from the band discontinuity across a heterojunction.[4, 5] Unlike the *pn* junction solar cell in which a built-in potential is needed to separate the photogenerated electron-hole pairs in the bulk, electrons and hole are spontaneously dissociated onto the two sides of the interface in an excitonic solar cell. The interfacial band alignment thus provides an exothermic pathway for exciton dissociation even though they do not possess sufficient energy to dissociate in the bulk. The huge chemical potential gradient created across the interface significantly increases the open-circuit voltage beyond the built-in potential.[6]

Recent advances of two-dimensional (2D) materials and their van der Waals heterostructures (vdWHs)[7-9] unravel an exciting route towards advanced excitonic solar cell technology.[10, 11] The vertical coupling of different 2D materials via weak van der Waals (vdW) forces results in '*designer*' heterostructures with physical properties not easily found in the nature, thus enabling myriads of high-performance nanoelectronics, optoelectronics, photonics and neuromorphic devices to be realized.[4, 5, 12-23] The optical absorption, band alignment and band gap of vdWHs can be custom made with ease. Together with the long exciton lifetime and the capability of efficient charge harvesting at ultrafast timescales[13-15], vdWH has become a compelling and potentially disruptive platform for the development of ultimately compact excitonic solar cell technology.[5, 24, 25] Designing high-performance vdWH excitonic solar cell is, however, a complex technological challenge both theoretically and experimentally. Computationally, the design space of vdWHs is enormously large due to nearly unlimited material combinations and the continual discovery of new 2D materials. The maximum power conversion efficiency (PCE) of vdWHs is predicted to lie within the range of 10% to 20%, with only a minority of vdWHs[20, 21, 26] (e.g. GaAs/InAs, $HfSe_2/GeO_2$ and BAs/BP) capable of achieving large PCE over 20%. Experimentally, the PCE of fabricated devices are only a small fraction of the theoretical limit[10] due to material defects and carrier recombination losses.[5] Furthermore, the experimentally realizable material combinations of vdWHs are constrained by the material synthesis and assembly processes.[27, 28] The limited availability of high-PCE vdWHs and the formidable challenges of fabricating high-quality vdWHs highlight the urgency and importance of continual computationally searching for high-PCE vdWH, so to further expand the ground for the experimental realization of high-performance excitonic solar cell technology.

In this work, we show that the vdWH family composed of septuple-layered $MoSi_2N_4$ and $WSi_2N_4$ monolayers[29] – the recently discovered archetypes of the emerging $MA_2Z_4$ monolayer family – offers a versatile material platform towards high-efficiency solar energy harvesting, optoelectronics and photonics applications. Using first-principle density functional theory (DFT) simulations, we constructed 51 types of $(Mo,W)Si_2N_4$-based vdWHs and characterized their electronic and optical properties. The vdWHs are systematically classified according to their contact types and band alignments.[30] Remarkably, we found that $MoSi_2N_4$/(ZnO, $WGe_2N_4$) and $MoSi_2N_4$/(InSe, $WSe_2$) are Type-II vdWHs with PCE reaching well over 15% and 20% respectively, which are comparable to many state-of-art solar cells composed of silicon and perovskite[31] as well as vdWHs.[20, 26] Furthermore, majority of the 51 types vdWHs exhibit substantial optical absorption in the visible and ultraviolet (UV) spectrum. Nearly all vdWHs exhibit peak optical absorption in the UV regime, thus suggesting their potential in UV photonics applications. Our work unravels the enormous potential of (Mo,W)$Si_2N_4$-based vdWHs in UV photonics, optoelectronics and solar energy harvesting applications, and call for a more comprehensive computational and experimental search of vdWHs with exceptional physical properties in the expansive $MA_2Z_4$ material family.

## 2. Computational Methods

QuantumATK[32] is used to construct the heterostructures with lattice mismatch of less than 5%[33, 34] between monolayers using supercell structure and applying the appropriate rotation. A vacuum layer of thickness ~25 Å is inserted between adjacent bilayer unit cell to eliminate the interactions from periodic images. We used the Vienna Ab initio Simulation Package (VASP)[35] for structural optimization, energy and charge calculation, and VASPKIT[36] for optical properties calculation. For structural optimization and self-consistent field calculations, a Gamma-centered Brillouin zone k-point sampling grid of 11 x 11 x 1 with Monkhorst-Pack method is used. In the structural optimization step, the supercell volume is kept constant while the cell shape and ion positions are allowed to change. The break condition for each ionic relaxation loop is set to $10^{-3}$ eV Å$^{-1}$ such that the forces acting on the unit cell after convergence is less than this value and the break condition for each electronic relaxation loop is set to $10^{-6}$ eV. To take into account the weak van der Waals interactions between the monolayers, we adopted the DFT-D3 method with the Grimme scheme. We chose the generalized gradient approximation (GGA) with the Perdew-Burke-Ernzerhof (PBE) form for the exchange correlation functional in the calculations for all our heterostructures, since $MoSi_2N_4$ band gap obtained from PBE (i.e. 1.73 eV and 1.78 eV)[37, 38] is closer to the experimental value of 1.94 eV,[29] as compared to the Heyd-Scuseria-Ernzerhof hybrid functional form (HSE06) with bandgap of 2.30 eV.[29] Moreover, it has been demonstrated that the band alignment types – one of the key results of this work – predicted using PBE is in good agreement with that obtained using HSE06.[30] The supercells of the vdWHs are shown in **Figures S1-S6**. The electronic band structures, plane-averaged electrostatic potential profiles, plane-averaged differential charge density profiles, dielectric constants, optical absorption spectra and the general calculation data of all 51 types of vdWHs are compiled and systematically presented in **Figures S7-S13**, **S14-S19**, **S20-S25**, **S26-S31**, **Tables S1** and **S2** of the **Supporting Information** respectively.

### 3. Results and Discussion

**3.1 $MoSi_2N_4$ and $WSi_2N_4$: An Overview of their van der Waals Heterostructures**

The discovery of $MoSi_2N_4$ and $WSi_2N_4$ open an exciting avenue towards the development of novel two-dimensional (2D) material device technology.[29] Monolayer $MoSi_2N_4$ and $WSi_2N_4$ are synthetic 2D semiconductor with exceptional electrical mobilities that outperforms the widely studied $MoS_2$ monolayer and hold great promises in nanoelectronics and sub-10-nm transistor applications.[39-41] Structurally, $MoSi_2N_4$ is a septuple-layered composed of a transition metal nitride sub-monolayer (Mo-N) intercalated between two outer silicon-nitride (Si-N) sub-monolayers. Such intercalated architecture gives rise to a large family of six-membered ring monolayers with septuple-atomic-layer structure, known as the $MA_2Z_4$ monolayer family.[42] Intriguingly, the band edge states of $MoSi_2N_4$ are localized around the inner Mo-N sublayer and are protected by the outer Si-N sublayer[37, 43] from external influences. Such unusual *built-in* atomic-layer protection mechanism renders $MoSi_2N_4$ as exceptional robust against environmental coupling effects and enables the formation of highly energy-efficient Ohmic contact without severe degradation of its electronic properties.[37] Recent studies have unravelled myriads of exceptional behaviours in $MoSi_2N_4$,[44-47] including strong exciton-phonon coupling at room temperature,[48] strain-tunable optical response,[49] tunable electronic properties via molecular doping,[50] field-induced metal-semiconductor transition,[44] large piezoelectric coefficient,[51] sizable thermoelectric figure of merit,[45] strained enhanced solar energy absorption[52] and ultrahigh thermal conductivities.[53] Beyond $MoSi_2N_4$ and $WSi_2N_4$ monolayers, the family of $MA_2Z_4$ monolayers[42] also exhibit enormously rich physical and chemical behaviours, including nontrivial topological phases,[42] valley-contrasting physics,[54-57] nonlinear optical response,[58] hydrogen evolution catalytic performance,[59] and superior electrocatalytic activity for energy storage application.[60]

Driven by the success of vdWH material engineering, $MoSi_2N_4$-based vdWHs have also been intensively studied recently. For example, $MoSi_2N_4$ interfaced with InSe,[61] blue phosphorus,[62] $C_2N$,[63] $CrCl_3$,[64] $C_3N_4$,[65] $MoS_2$,[66] graphene,[38] $NbS_2$,[38] GaN,[67] ZnO,[67] $MoSe_2$,[68] $Cs_3Bi_2I_9$,[69] $MoGe_2N_4$,[70] $MoGeSiN_4$,[71] and MoSH[72] have been computationally predicted to exhibit efficient photocatalytic water splitting,[61-63] strain-tunable valley splitting effect,[64] strong optical absorption in the visible light regime with strain and field-effect tunable band structures.[38, 65-73] Nonetheless, the *catalogue* of $(Mo,W)Si_2N_4$-based vdWHs remains largely incomplete thus far.

As summarized in **Figure 1**, 51 types of $(Mo,W)Si_2N_4$-based vdWHs are constructed in this work, which are composed of $(Mo,W)Si_2N_4$ monolayers in contact with seven classes of 2D materials [see **Figure 1(A)**]: (i) semiconducting transition metal dichalcogenides (TMDC), such as $MoS_2$ and $WS_2$; (ii) Janus semiconducting TMDC, such as MoSSe and WSSe; (iii) metallic TMDC, such as $NbS_2$ and $1T-MoS_2$; (iv) Honeycomb monolayers, such as the semimetallic graphene and insulating hBN; (v) Group III-IV monolayer of InSe; (vi) ultrathin topological Dirac semimetals of $Na_3Bi$ monolayer; and (vii) other members from the $MA_2Z_4$ family, such as $\alpha_2$-MoGe$_2$P$_4$, MoGe$_2$N$_4$, as well as the vdWH formed by the mutual stack of MoSi$_2$N$_4$/WSi$_2$N$_4$. In **Figure 1(B)**, the 51 vdWHs are systematically categorized into six sub-types, namely: (i) Metal/Semiconductor (MS) Ohmic contact (4 species); (ii) MS Schottky contact (6 species); and Semiconductor/Semiconductor (SS) contact with (iii) Type-I (19 species); (iv) Type-II (15 species); (v) Type-III (2 species), and (vi) unconventional hybridized-type (5 species) band alignments [see **Figure 1(C)** for schematic drawings of the band diagram of different contact types]. The band alignment diagrams of the 25 types of 2D monolayers considered in this work are shown in **Figure 2(A)** which provides a qualitative understanding on the contact type and band alignment formation via Anderson rule (AR).[74] Due to the finite interaction between the two constituent monolayers,[75] AR may no longer be valid in vdWHs with sizable interlayer coupling and interface dipole.[76, 77] As shown in **Figure 2(B)**, AR remains valid for most of the metallic TMDC, graphene, InSe, hBN and MA$_2$Z$_4$ family due to their weak coupling nature. In contrast, the disagreement between AR and direct DFT calculations is rather severe for semiconducting and Janus TMDC monolayers as well as the honeycomb semiconducting monolayers of GaN, BP and ZnO. This disagreement highlights the importance of direct DFT calculations for contact type and band alignment determination.

### 3.2 Metal/Semiconductor (MS) Contacts

We first consider (Mo,W)Si$_2$N$_4$-based vdWHs of MS contact type. In this case, either Ohmic or Schottky contacts can be formed at the interface.[78, 79] For an Ohmic contact, the absence of a Schottky barrier greatly facilitate charge injection across an MS contact and is much sought after for creating energy-efficient electrical contacts.[80] MS vdWH contacts are particularly attractive due to the strong suppression of the Fermi level pinning effect, a detrimental effect that severely impedes the engineering of energy-efficient electrical contacts.[81, 82] Here we identify 4 species of *p*-type Ohmic contacts formed by MoSi$_2$N$_4$ and WSi$_2$N$_4$ in contact with NbS$_2$ and NbSe$_2$ [see **Figure 3(A-D)**]. The formation of these *p*-type Ohmic vdWHs can be readily understood from the band alignment diagram in **Figure 2(A)**, where the relatively larger work functions of NbS$_2$ (6.06 eV) and NbSe$_2$ (5.35 eV) align with the valence band maximum (VBM) energies of MoSi$_2$N$_4$ (5.30 eV) and WSi$_2$N$_4$ (5.14 eV) to yield *p*-type Ohmic contact by virtue of AR. We note that the MoSi$_2$N$_4$/NbS$_2$ contact is predicted to be an *p*-type Schottky contact with ultralow Schottky barrier height (SBH)[38] where the strain arising from lattice mismatch upon forming the vdWH is entirely distributed on the NbS$_2$ monolayer only. In contrary, strain is distributed evenly between the two monolayers in our calculations, which yields a *p*-type Ohmic contact instead.

For a bilayer vdWH made up of monolayer A and monolayer B, the plane-averaged differential charge density ($\Delta\rho$) is calculated as, $\Delta\rho = \rho_{AB} - \rho_A - \rho_B$, where $\rho_{AB}$ is the charge density of the heterostructure AB, $\rho_A$ is the charge density of monolayer A, $\rho_B$ is the charge density of monolayer B. The charge redistribution of MoSi$_2$N$_4$ and WSi$_2$N$_4$ contacted by NbS$_2$ and NbSe$_2$ are examined from the $\Delta\rho$ plots in the upper panel of **Figure 3(E-H)**. Significant electrons redistribution from the (Mo,W)Si$_2$N$_4$ monolayer towards the Nb(S$_2$,Se$_2$) monolayer occurs due to the higher VBM energies of the semiconducting monolayers when compared to the Fermi levels of the metallic monolayers. Such electron migration causes the *p*-doped characteristics of the (Mo,W)Si$_2$N$_4$ sub-monolayers upon contact formation. We further note that the interfacial charge redistribution occurs universally for all vdWHs as evident from the differential charge density plots in **Figures S20-S25**. The interfacial charge redistribution creates an interface dipole potential ($\Delta\Phi$) that manifests as an electrostatic potential energy difference between the two sides of the vdWHs. The $\Delta\Phi$ are marked directly in the plane-averaged electrostatic profile plots in **Figures S14-S19** for all vdWHs studied in this work.

Although SBH is absent in Ohmic vdWH contact, a van der Waals gap is present between the two contacting monolayers that may impede charge ejection efficiency.[83] The tunneling potential barrier $\Phi_{TB}$ and the barrier width $d_{TB}$ can be extracted from the plane-averaged electrostatic potential plots with reference to the Fermi level [see bottom panel of **Figure 3(E-H)**]. For MoSi$_2$N$_4$/NbS$_2$, MoSi$_2$N$_4$/NbSe$_2$, WSi$_2$N$_4$/NbS$_2$ and WSi$_2$N$_4$/NbSe$_2$, the tunneling potential barriers are 5.14, 4.70, 5.04, 4.60 eV respectively, and the barrier widths are 1.88, 1.86, 1.85, 1.84 Å respectively. Accordingly, the charge injection efficiency across the van der Waals gap can then be characterized by the tunneling-specific resistivity ($\rho_t$) using Simmons formula:[84]

$$\rho_t = \frac{8\pi^2 \hbar^2 d_{TB}}{3e^2 \sqrt{2m_e \Phi_{TB}}} \exp\left(\frac{2d_{TB}\sqrt{2m_e \Phi_{TB}}}{\hbar}\right)$$

**(1)**

where $\hbar$, $e$, $m_e$ is the Planck's constant, charge magnitude and mass of the free electron respectively. The calculated $\rho_t$ lies in the order of 10$^{-9}$ which are comparable to those of metal/MoSi$_2$N$_4$ and metal/WSi$_2$N$_4$ contacts (i.e. from $0.1 \times 10^{-9}$ to $3.85 \times 10^{-9}$ $\Omega cm^2$)[37] and that in the semimetal contact of Bi/MoS$_2$ contact ($1.81 \times 10^{-9}$ $\Omega cm^2$).[85] Finally, when (Mo,W)Si$_2$N$_4$ monolayers are interfaced with (Mo,W)S$_2$ in metallic 1T-phase and with graphene, the resulting (Mo,W)Si$_2$N$_4$/(1T-MoS$_2$, 1T-WS$_2$, graphene) vdWHs form *p*-type Schottky contacts [see **Figure 4(B)**]. For Schottky type vdWHs, the charge transfer follows the same trend of electron migration from (Mo,W)Si$_2$N$_4$ towards the metallic monolayers, but the overall plane-averaged magnitude of $\Delta\rho$ are far lesser than that of the Ohmic counterparts [see **Figure S20 and S21**]. The resulting *p*-type SBHs of these Schottky vdWHs are extracted from the electronic band structure as the energy difference between the Fermi level of the metallic monolayer and the VBM of the semiconducting monolayer.[38] The SBHs of these MS vdWHs cover a broad energy range between 0.3 eV and 0.9 eV [see **Figure 4(C)**], which can be useful for Schottky diode and current rectification applications.

**3.3 Semiconductor/Semiconductor (SS) Contacts**

We classify all SS vdWHs according to the relative location of the conduction band minimum (CBM) and VBM in the energy space, i.e. the three conventional band alignment types:[86-88] (i) Type-I straddling gap band alignment where the CBM and VBM arises dominantly from only one of the constituent monolayers; (ii) Type-II staggered gap band alignment where the CBM and VBM resides dominantly in different sub-monolayers; and (iii) Type-III broken gap band alignment where the CBM of one sub-monolayer overlaps with the VBM of another sub-monolayer. Unconventional band alignment types can also occur in vdWH in which the CBM and VBM energies cannot be straightforwardly classified according to these conventional band/or alignment types (denoted as Type-H in this work).[89] In this Section, the 41 types of SS (Mo,W)Si$_2$N$_4$-based vdWHs are classified according to their band alignment types [see **Figure 5(A)**]. The electronic band structures of representative Type-I, II, III and H vdWHs are shown in **Figure 5(B-E)** respectively.

### 3.3.1 Band Alignment Types

The band gap values, the direct or indirect nature of the band gap, and the band alignment types (i.e. Type-I, II and H) are summarized in **Figure 5(A)**. We identify 19 species of Type-I vdWHs – a band alignment type that is useful for photoluminescence applications, such as light-emitting diodes and lasers, due to the localization of CBM and VBM states within the same sub-monolayer. Among these Type-I vdWHs, 15 species possess direct band gap that is highly beneficial for engineering efficient excited electron-hole pair recombination.[90] The electronic band structure of a representative Type-I vdWH, MoSi$_2$N$_4$/hBN is shown in **Figure 5(B)**. We note that the MoSi$_2$N$_4$/MoS$_2$ calculated using PBE (HSE06) here shows a direct band gap of 0.60 eV (1.06 eV) at the Γ point which is in stark contrast to the 1.26 eV (1.84 eV) reported in a previous work.[66] We attribute this difference to the stacking orientation of the vdWH construction. The biaxial strain arising from lattice mismatch of the supercell considered in this work is only 2.7% (i.e. 30° rotation between the two monolayers), as compared to the substantially larger biaxial strain of 7.8% in the previous work.[66]

For Type-II band alignment, 15 species of vdWHs are identified. **Figure 5(C)** shows the electronic band structure of $MoSi_2N_4/WSi_2N_4$, a Type-II vdWH compose entirely of $MA_2Z_4$ monolayers. The Type-II band alignment and the indirect nature of $MoSi_2N_4/WSi_2N_4$ is in good agreement with a recent DFT calculation of the same vdWH but under different stacking configuration.[91] Type-II vdWHs are potential candidates for excitonic solar cell application. The maximum PCEs calculated for each Type-II vdWHs are presented in Section 5.2 below. Finally, we note that $(Mo,W)Si_2N_4/Na_3Bi$ are the only Type-III vdWHs. Such band alignment arises from the ultralow work function[92] and the narrow band gap[39] nature of $Na_3Bi$ monolayer [see **Figure 2(A)**]. The VBM of $Na_3Bi$ overlap strongly with the CBM of $(Mo,W)Si_2N_4$, thus leading to the formation of Type-III SS contacts. **Figure 5(D)** shows the electronic band structure of $WSi_2N_4/Na_3Bi$ as an illustrative example.

Beyond the conventional band alignment types discussed above, we further observed 5 vdWHs with unconventional band alignments, namely the $(Mo,W)Si_2N_4/MoS_2$, $MoSi_2N_4/SMoSe$, $MoSi_2N_4/\alpha_2\text{-}WGe_2P_4$ and $WSi_2N_4/SWSe$ [see **Figure 5(E)** for the electronic band structure of a representative vdWH, $WSi_2N_4/MoS_2$]. As the VBM electronic states are delocalized and distributed rather evenly among the two sub-monolayers, the conventional Type-(I, II ,III) classification scheme is no longer applicable and we denote these vdWHs as the *hybridized* type (denoted as Type-H), which are intermediate between Type-I and Type-II vdWHs due to the spatial distribution of VBM states in both sub-monolayers. The Type-H contacts are also observed in vdWHs composed of TMDC monolayers[30, 93] and binary semiconducting monolayers composed of group IV–V elements[89] which arise due to the orbital hybridization of the out-of-plane $p_z$-states. As the VBM states are distributed among the two sub-monolayers in Type-H contacts, some of the excited electron-hole pairs can recombine within the same sub-monolayers, thus rendering charge separation much less effective and may impede the potential of Type-H vdWHs for optoelectronics applications. We further note that to verify these unusual band structures, we employ the HSE06 exchange correlation functional in the electronic band structure calculations of these heterostructures. The Gamma-centered Brillouin zone k-point sampling grid of 6 x 6 x 1 is used due to computational resources constraint. Interestingly, the VBM hybridization is removed in $MoSi_2N_4/2H$-$MoS_2$ and $MoSi_2N_4/\alpha_2$-$WGe_2P_4$ under the HSE06 calculations [see **Figure S13**].

### 3.3.2 Excitonic Solar Cell Maximum Conversion Efficiency of Type-II vdWHs

Optically excited electron-hole pairs are spatially well-separated in Type-II vdWHs, which enables the efficient dissociation of these carriers to form conduction current. Such vdWHs are thus practically useful for applications where long-live and/or delocalized electron-hole pairs are needed, such as photodetector, excitonic solar cell and photocatalytic water splitting.[87] We calculated the maximum solar cell PCE of these Type-II vdWHs, which represents the theoretical upper limit of solar-to-electricity conversion in a heterostructure device. The PCE is calculated as:[86, 94]

$$\eta = \frac{0.65\left(E_g^d - \Delta E_c - 0.3\right) \int_{E_g^d}^{\infty} \frac{P(\hbar\omega)}{\hbar\omega} d(\hbar\omega)}{\int_0^{\infty} P(\hbar\omega)\, d(\hbar\omega)}$$

(2)

where $E_g^d$ is the bandgap of the donor monolayer, $\Delta E_c$ is the conduction band offset (CBO) between the two monolayers, $P(\hbar\omega)$ is the solar spectral irradiance as a function of photon frequency $\omega$, and the 0.3 eV term in **Equation (2)** is an empirical parameter to account for the energy loss during the conversion process.[95] The AM1.5G standard solar spectrum is used for the solar spectral irradiance $P(\hbar\omega)$, which yields an integrated solar intensity of $P_{solar} = \int_0^{\infty} P(\hbar\omega)\, d(\hbar\omega) \approx 1000$ W/m². A representative value of 0.65 is assumed for the electrical fill factor of the solar cell[95] which corresponds to the typical situation where ~60% of the excited electron-hole pair are dissociated to form the short-circuit current as observed experimentally.[96] To achieve high PCE, the donor monolayer band gap should be within the desirable range of 1.0 to 1.7 eV, while the CBO should be as small as possible so to suppress the energy loss.

The calculated PCEs are listed in **Figure 6(A) and 6(B)** for $MoSi_2N_4$-and $WSi_2N_4$-based Type-II vdWHs, except for $WSi_2N_4/MoGe_2N_4$ in which the overly small band gap of $E_g - \Delta E_c < 0.3$ is unsuitable for solar cell application. In general, $MoSi_2N_4$-based Type-II SS contacts exhibit higher overall PCE when compared to the $WSi_2N_4$-based Type-II vdWHs. Recent high throughput screening of 581 Type-II vdWHs[26] based on Anderson rule has uncovered 17 candidate vdWHs with exceptional PCE higher than 20%, and the highest being that of $HfSe_2/GeO_2$ (PCE = 22.6%). In our case, the PCE of $MoSi_2N_4/InSe$ and $MoSi_2N_4/WSe_2$ vdWHs reaches 20.5% and 22.0% respectively, which is comparable to that of $HfSe_2/GeO_2$ and higher than many other vdWHs, such as phorphorus-based vdWHs (e.g. $MoS_2$/Phosphorene with PCE of 17.5%).[97] The exceedingly large PCE of $MoSi_2N_4/InSe$ and $MoSi_2N_4/WSe_2$ can be attributed to the small CBO of 0.03 eV and 0.05 eV respectively. Furthermore, $MoSi_2N_4/InSe$ and $MoSi_2N_4/WSe_2$ are indirect band gap Type-II heterostructures in which the recombination of excitons is expected to be less effective due to the need of phonon-assisted process to compensate the momentum-mismatch between CBM and the VBM states.[98] $MoSi_2N_4/InSe$ and $MoSi_2N_4/WSe_2$ are thus expected to host long-live interlayer exciton that are highly beneficial for highly efficient light-to-electricity conversion.

For the four $MoSi_2N_4$-based Type-II vdWHs with PCE over 15%, we calculated the real and imaginary parts of their dielectric function as well as the optical absorption spectra [see **Figure 6(C-F)**]. All four heterostructures exhibit substantial optical absorption of >10% in the visible light regime [denoted by dashed vertical lines in **Figure 6(C-F)**]. Particularly, the vdWH with the highest PCE, i.e. MoSi$_2$N$_4$/WSe$_2$ exhibits a strong optical absorption shoulder (>20%) in the visible light regime of energies 2.5 eV and above, thus suggesting their potential usefulness for solar energy absorption and photodetection applications.[99-101] Comparing to the isolated MoSi$_2$N$_4$ and WSi$_2$N$_4$ monolayers [see **Figure 6(G) and 6(H)**], the overall optical absorption in the visible light regime is significantly enhanced upon forming the vdWHs, thus establishing the importance of vdW engineering in boosting the solar energy conversion capability of MoSi$_2$N$_4$ and WSi$_2$N$_4$.

**3.4 Optical Properties**

The optical properties of the 51 vdWHs are calculated based on quantum mechanical interband transition process from the valence bands to the conduction bands. The macroscopic frequency-dependent complex dielectric function $\varepsilon(\omega)$ is calculated based on the random phase approximation formalism (RPA).[36] The optical absorbance ($\alpha$) of a free-standing 2D monolayer under normal illumination can then be calculated as:[102]

$$\alpha = \frac{Re(\tilde{\sigma}(\omega))}{\left|1 + \frac{\tilde{\sigma}(\omega)}{2}\right|^2}$$

(3)

where $\tilde{\sigma}(\omega) \equiv \sigma_{2D}(\omega)/\varepsilon_0 c$, $\sigma_{2D}(\omega) = i\varepsilon_0 \omega L(1 - \varepsilon(\omega))$ is the frequency-dependent complex 2D optical conductivity, $\varepsilon_0$ is the vacuum permittivity, $L$ is the length of the in-plane lattice constant, $c$ is the speed of light in vacuum. **Equation (3)** yields a maximum absorbance of $\alpha_{max} = 0.5$ when $Re[\tilde{\sigma}(\omega)] = 2$ and $Im[\tilde{\sigma}(\omega)] = 0$, which represents the theoretical absorption limit of a free-standing monolayer.[102]

The complex dielectric function and optical absorbance of the 51 types of vdWHs are calculated and presented in **Figures S26-S31**. In general, the (Mo,W)Si$_2$N$_4$ vdWH family exhibits sizable optical absorption in the visible regime with optical absorbance up to 25%.

We compiled the peak absorbance for photon energies up to 5 eV for all vdWHs as shown in **Figure 7**. Intriguingly, almost all vdWHs exhibit peak absorbance in the UV regime, except for $MoSi_2N_4$/2H-$WS_2$ vdWH which has a peak absorbance at the visible photon energy of 3.2 eV. The peak absorbance of the isolated $MoSi_2N_4$ and $WSi_2N_4$ monolayers [see **Figure 6(G) and 6(H)** for the optical absorption spectra of $MoSi_2N_4$ and $WSi_2N_4$ monolayers] is 20% at 4.0 eV and 17% at 4.4 eV respectively, which are comparable to some of the strongest mid-UV-absorbing 2D materials (e.g. $SnS_2$ and $SnSe_2$).[102] More importantly, the peak absorbance is significantly enhanced upon forming the vdWHs, with many vdWHs reaching a peak absorbance of over 30% in the UV regime. In general, the $WSi_2N_4$ vdWHs exhibit peak absorbance at higher photon energies as compared to that of $MoSi_2N_4$ counterpart in the energy range being studied, owing to the slightly higher peak absorbance photon energies of 4.4 eV of the $WSi_2N_4$ monolayer, as compared to that of monolayer $MoSi_2N_4$ at 4.0 eV [see **Figure 6(G) and 6(H)**]. We grouped the vdWHs into the sub-groups of near-UV (3.5 eV to 4.0 eV), mid-UV (4.1 eV to 4.4 eV) and deep-UV (>4.5 eV) according to the peak absorbance photon energies. Intriguingly for the vdWHs composed of entirely $MA_2Z_4$, (i.e. $(Mo,W)Si_2N_4/\alpha_2$-$(Mo,W)Ge_2P_4$) the vdWHs exhibit exceptionally strong UV absorption of over 40%, which approaches the optical absorption limit of a free-standing monolayer, and a broadband strong absorption shoulder between 3 eV and 5 eV [see **Figure S29 and S31**]. These findings suggest the potential strength of $MA_2Z_4$-based vdWHs in photonics device applications.

## 4. Conclusion

In summary, we investigated 51 types of $MoSi_2N_4$ and $WSi_2N_4$ vdWHs and categorized these heterostructures based on their contact and band alignment types. For MS Ohmic contacts, the tunnelling-specific resistivities are comparable or lower than many of the previously reported 2D semiconductor electrical contacts. Multiple Type-II heterostructures are identified. Particularly, $MoSi_2N_4$/InSe and $MoSi_2N_4$/$WSe_2$ exhibit excellent solar cell energy conversion efficiency greater than 20% and substantial optical absorption in the visible light regime, making them potential candidates for photovoltaic applications. In terms of optical absorption, the family of (Mo,W)Si$_2$N$_4$-based bilayer van der Waals heterostructures generally exhibit much enhanced ultraviolet optical absorption when compared to their isolated counterparts. The (Mo,W)Si$_2$N$_4$/($\alpha_2$-MoGe$_2$P$_4$,$\alpha_2$-WGe$_2$P$_4$) heterostructures have an exceedingly high ultraviolet peak absorbance of over 40% that approaches the theoretical maximum absorption limit of a free-standing monolayer, thus revealing their strength in ultraviolet photonics applications. Our findings reveal the promising role of (Mo,W)Si$_2$N$_4$-based van der Waals heterostructures as a versatile platform for designing novel optoelectronics and photonics applications, and call for a more comprehensive screening of the expansive MA$_2$Z$_4$ vdWH family to fully unlock their potential as building blocks of high-performance nanodevice technology.

**Supporting Information**
Supporting Information is available from the Wiley Online Library.

**Acknowledgements**
This work is supported by the Singapore Ministry of Education (MOE) Academic Research Fund (AcRF) Tier 2 grant (Grant No. MOE-T2EP50221-0019) and SUTD-ZJU IDEA Collaboration (Grant No. SUTD-ZJU (VP) 202001). G.W. is supported by the Science and Technology Research Program of Chongqing Municipal Education Commission (Grant No. KJQN202001402). C.C.T. is supported by the Singapore University of Technology and Design (SUTD) President's Graduate Fellowship (PGF) scholarship. L.C. is supported by the National Natural Science Foundation of China (Grant No. 12104136). S.-J.L. is supported by the National Natural Science Foundation of China (Grant No. 61974176). We thank Li Ling Tan and Jin Quan Ng for their assistance in data generation and analysis.

**Conflict of Interest**

The authors declare no conflict of interest.

# Figures

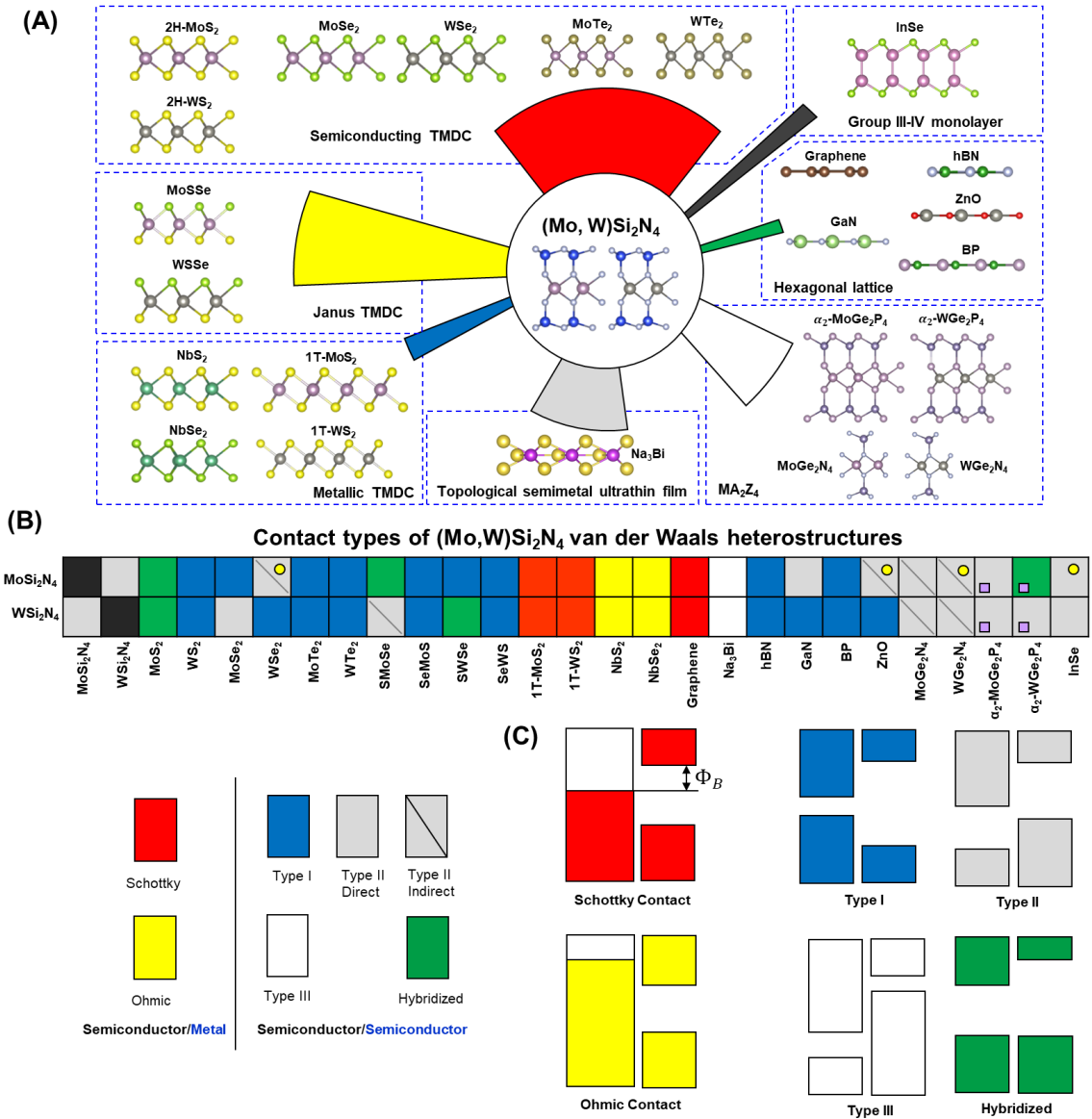

**FIGURE 1 – Classification of the 51 types of (Mo,W)Si₂N₄-based van der Waals heterostructures (vdWHs).** (A) Seven representative 2D material classes are considered, i.e. semiconducting, metallic and Janus TMDC, group III-IV monolayer (InSe), honeycomb lattices, ultrathin topological semimetal (Na$_3$Bi) and MA$_2$Z$_4$ monolayers. (B) Classification table of various vdWHs according to their contact types. Metal/semiconductor (MS) contacts are composed of Ohmic and Schottky contacts, while semiconductor/semiconductor (SS) contacts can be further classified into Type-I, Type-II (direct or indirect band gap), Type-III and the hybridized type in which the VBM of the two constituent monolayers overlap. The open circle represents vdWHs with solar cell maximum power conversion efficiency (PCE) higher than 15%. The open squares represent vdWHs with peak absorbance higher than 40%. (C) The band diagrams of different band alignment types.
28

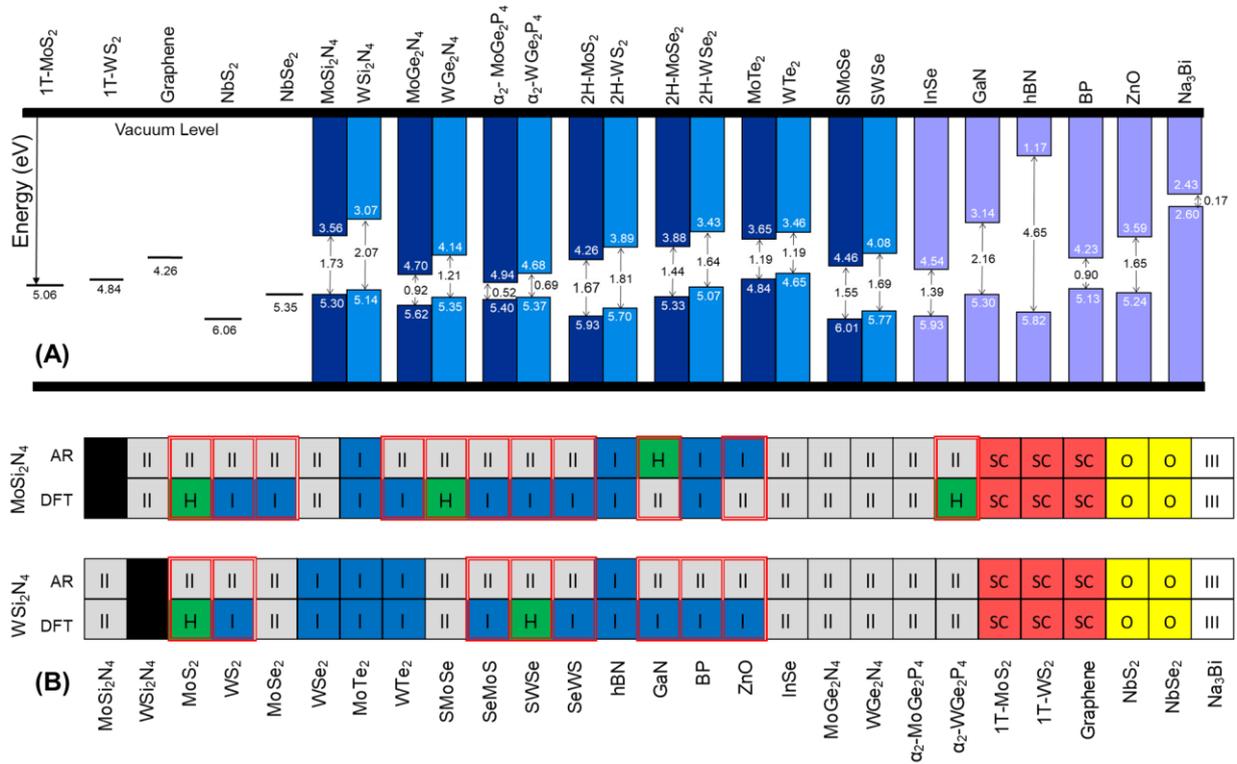

**FIGURE 2 – Band alignment of 2D monolayers and the assessment of Anderson rule.** (A) The CBM and VBM energies of the isolated 2D semiconducting monolayers, and the Fermi levels of 2D metallic and semimetallic monolayers. The energies are referenced to the vacuum level. (B) Side-by-side comparison of the band alignment and contact types between Anderson rule (AR) and direct DFT calculations (DFT). The band alignment types are indicated as 'I', 'II', 'III' and 'H' for Type-I, II, III and H vdWHs. The Schottky and Ohmic contacts are labelled as 'SC' and 'O' respectively. vdWHs with disagreement between AR and direct DFT calculation are highlighted by red bold frame.

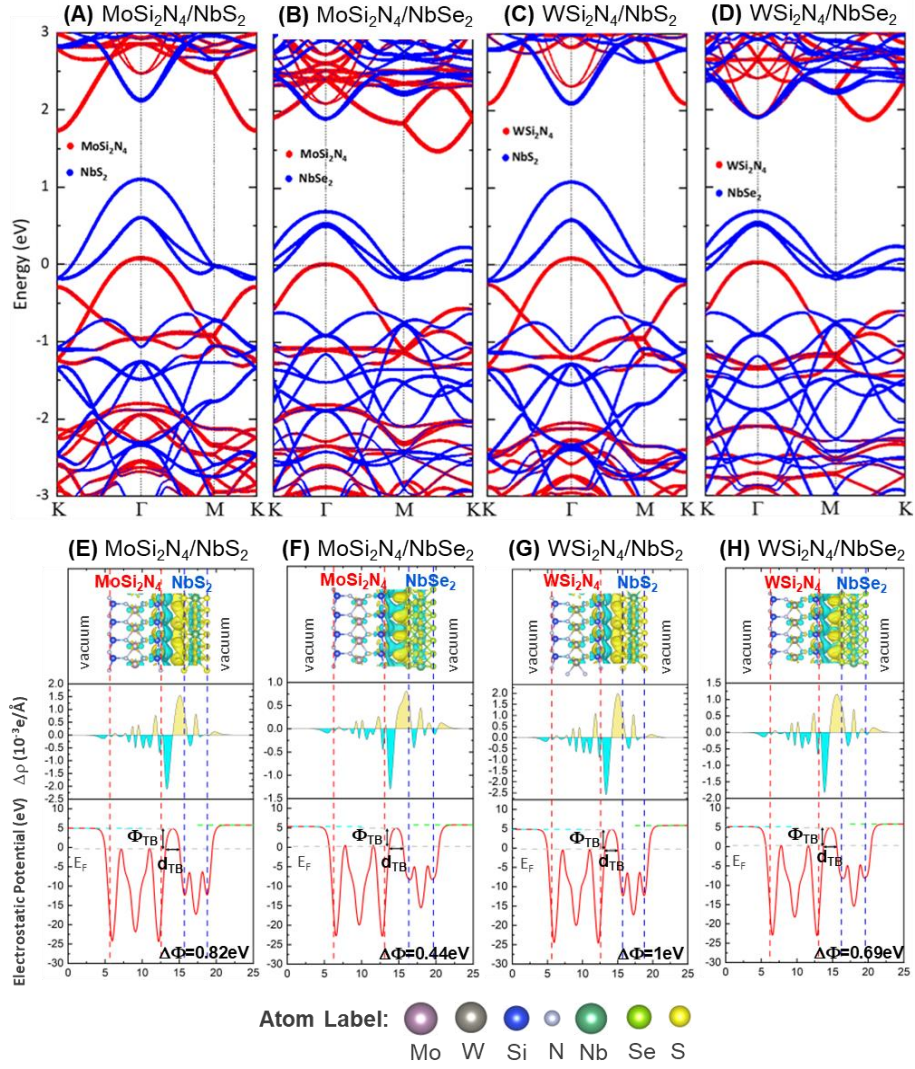

**FIGURE 3 – Electronic band structures, differential charge densities and electrostatic potential profiles of (Mo,W)Si$_2$N$_4$ Ohmic vdWHs.** (A) to (D) show the electronic band structures of MoSi$_2$N$_4$/NbS$_2$, MoSi$_2$N$_4$/NbSe$_2$, WSi$_2$N$_4$/NbS$_2$ and WSi$_2$N$_4$/NbSe$_2$ respectively. (E) to (H) show the plane-averaged differential charge density (upper and middle panels) and the electrostatic potential profile (bottom panel) of the Ohmic contact of (A) to (D) respectively. The yellow(cyan)-shaded region represents electron accumulation (depletion).

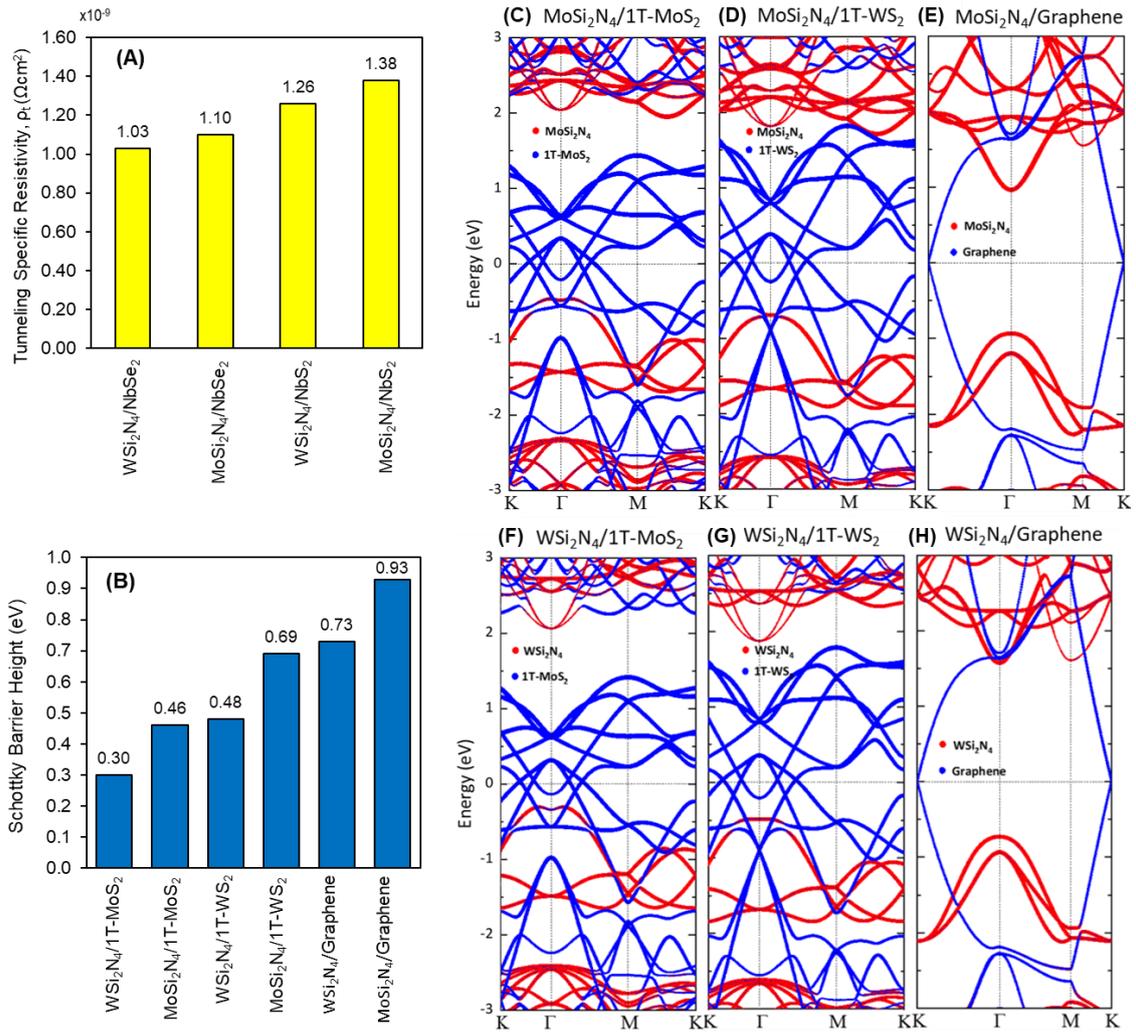

**FIGURE 4 – Tunneling-specific resisitivity of Ohmic MS contacts and Schottky barrier heights of Schottky MS contacts.** (A) The tunneling-specific resistivity for 4 types of Ohmic MS contact; and (B) the Schottky barrier heights are calculated for the p-type Schottky vdWHs respectively. The electronic band structures of the p-type Schottky vdWHs are shown in (C) to (H).

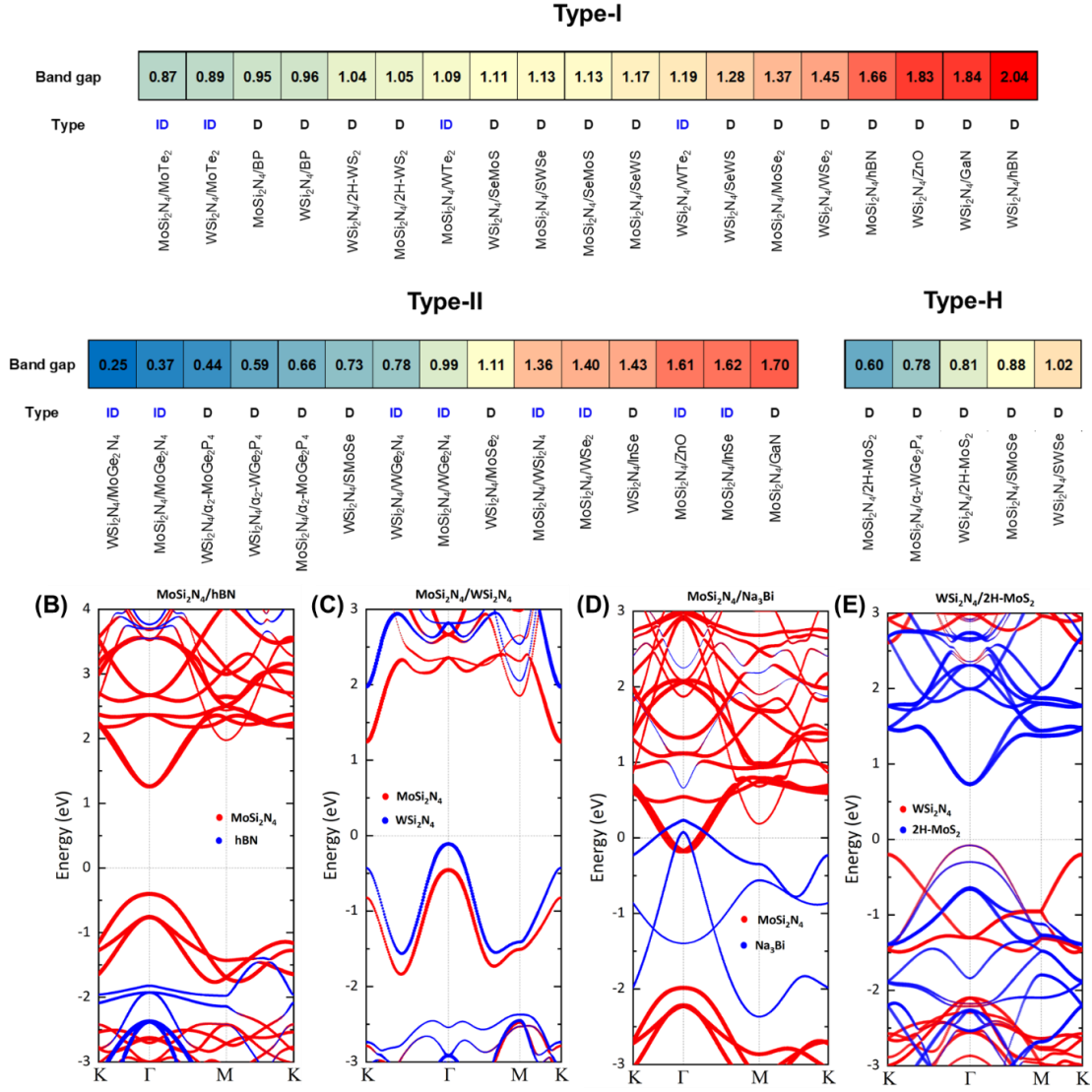

**FIGURE 5** – **Band alignment types, band gap, and the direct or indirect nature of the band gap, and electronic band structures for Type-I, II and H vdWHs.** (A) The band gap values are quoted in the unit of eV. The symbols "D" and "ID" denote direct and indirect type of the band gap, respectively. Electronic band structures of representative Type-I, II, III and H vdWHs of (B) $MoSi_2N_4$/hBN, (C) $MoSi_2N_4$/$WSi_2N_4$, (D) $MoSi_2N_4$/$Na_3Bi$, and (E) $WSi_2N_4$/$MoS_2$ respectively.

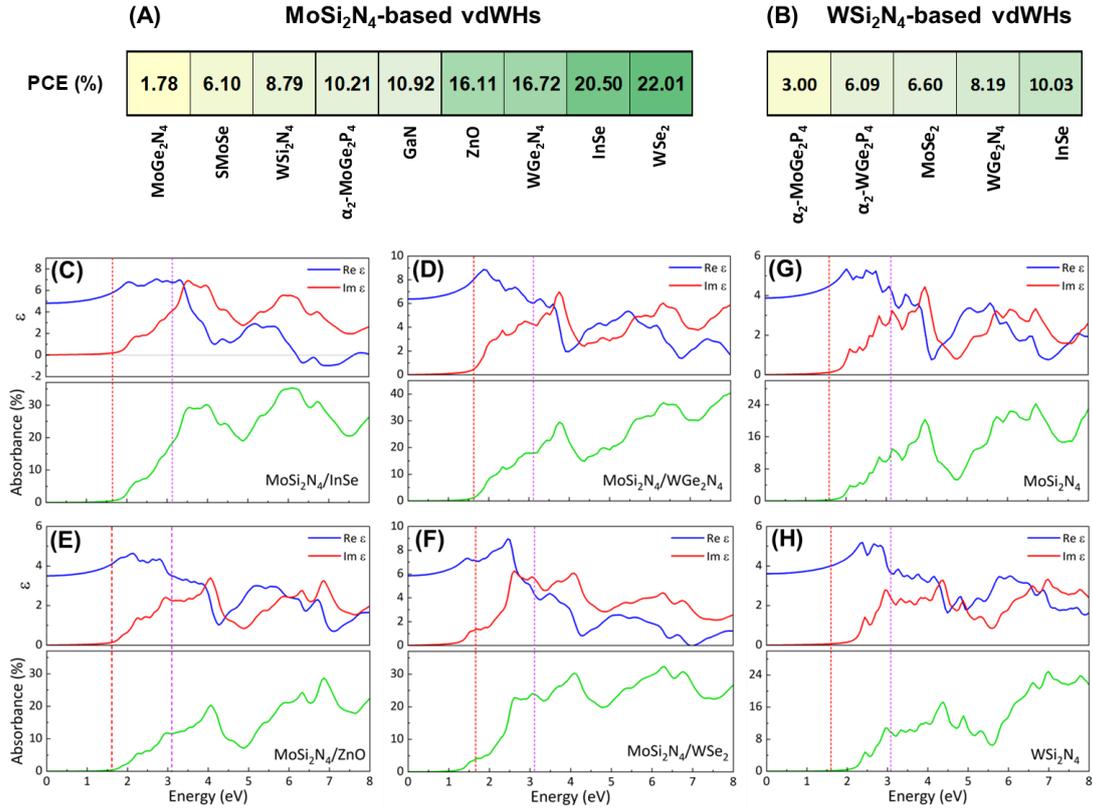

**FIGURE 6** – **Solar cell maximum power conversion efficiency (PCE) of Type-II SS vdWHs.** (A) and (B) shows the calculated PCE for $MoSi_2N_4$ and $WSi_2N_4$ vdWHs respectively. The dielectric constants and the optical absorption spectra of Type-II SS vdWHs with PCE higher than 15%: (C) $MoSi_2N_4$/ZnO, (D) $MoSi_2N_4$/$WGe_2N_4$, (E) $MoSi_2N_4$/InSe, and (F) $MoSi_2N_4$/$WSe_2$. The dielectric constant and optical absorption spectra of isolated $(Mo,W)Si_2N_4$ are shown in (G) and (H) respectively.

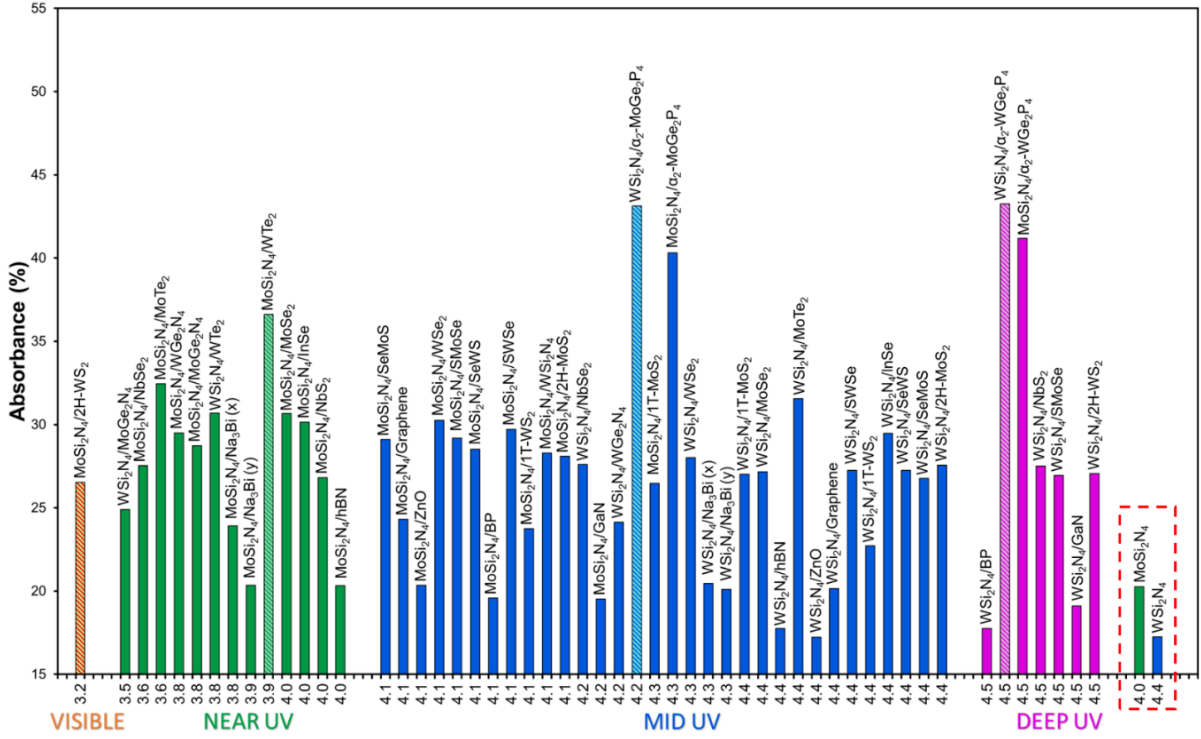

**FIGURE 7 – Peak absorbance of 51 species of (Mo,W)Si$_2$N$_4$-based vdWHs.** The vdW with the largest maximum absorbance in the near-UV (3.5 eV to 4.0 eV), mid-UV (4.1 eV to 4.4 eV) and deep-UV (> 4.5 eV) regimes are shaded. The dashed box denote the peak absorbance of the isolated (Mo,W)Si$_2$N$_4$ monolayers. (Mo,W)Si$_2$N$_4$/Na$_3$Bi exhibits in-plane optical dichroism as labelled by the (x) and (y) orthogonal directions.